# Ties That Bind - Characterizing Classes by Attributes and Social Ties

Aria Rezaei
Stony Brook University
Stony Brook, NY
arezaei@cs.stonybrook.edu

Bryan Perozzi[*]
Google Research
New York, NY
bperozzi@acm.org

Leman Akoglu
H. John Heinz III College
Carnegie Mellon University
lakoglu@cs.cmu.edu

## ABSTRACT

Given a set of attributed *subgraphs* known to be from different *classes*, how can we discover their *differences*? There are many cases where collections of subgraphs may be contrasted against each other. For example, they may be assigned ground truth labels (spam/not-spam), or it may be desired to directly compare the biological networks of different species or compound networks of different chemicals.

In this work we introduce the problem of characterizing the differences between attributed subgraphs that belong to different classes. We define this characterization problem as one of partitioning the attributes into as many groups as the number of classes, while maximizing the total attributed quality score of all the given subgraphs.

We show that our attribute-to-class assignment problem is NP-hard and an optimal $(1 − 1/e)$-approximation algorithm exists. We also propose two different faster heuristics that are linear-time in the number of attributes and subgraphs. Unlike previous work where only attributes were taken into account for characterization, here we exploit *both* attributes and social ties (i.e. graph structure).

Through extensive experiments, we compare our proposed algorithms, show findings that agree with human intuition on datasets from Amazon co-purchases, Congressional bill sponsorships and DBLP co-authorships. We also show that our approach of *characterizing* subgraphs is better suited for sense-making than *discriminating* classification approaches.

## 1 INTRODUCTION

Besides connectivity, many graphs contain a state (or content) vector for each node. This type of graph is known as an *attributed* graph, and is a natural abstraction for many applications. For example, in a social network the profile information of individuals (e.g. age, occupation, etc.) constitute the attribute vector for each node. Biological data can also be represented as attributed graphs; protein-protein interaction (PPI) networks can have gene encodings of proteins as attributes, or gene interaction networks may contain gene ontology properties as attributes [14, 22].

We consider the following question: Given a collection of *attributed subgraphs* from different *classes*, how can we discover the attributes that characterize their *differences*? This is a general question, which finds applications in various settings depending on how 'subgraphs' and 'classes' are defined and interpreted. In social networks, subgraphs could be the local communities around each individual. That is because one's acquaintances carry a lot of information about them due to the factors of homophily (phenomenon that "birds of a feather flock together") and influence (phenomenon that our attitudes and behaviors are shaped by our peers) [7]. One can also consider the subgraphs extracted by a community detection algorithm, the social circles as defined by individuals, or any collection of small graphlets that come from an application (e.g. PPI networks of a collection of fly species). On the other hand, a 'class' corresponds to a broad categorization of subjects. In social network analysis, one may try to understand the differences between individuals living in different countries (e.g. U.S. vs. China), or having different demographics (e.g. elderly vs. teenagers). In biology, one may want to analyze PPI networks of healthy versus sick individuals or of mice versus humans.[1]

In this work, we propose to characterize the different classes *through the attributes that their subgraphs focus on*. Intuitively, we assume the nodes in each subgraph share a subset of attributes in common (e.g. a circle of friends who go to the same school and play baseball). That is, members of a subgraph "click" together through a shared characterizing attribute subspace, called the *focus attributes* [28]. It is expected that out of a large number of attributes, only a few of them would be relevant for each subgraph.

Our main insight for comparing subgraphs is then that the *subgraphs from different classes would exhibit different focus attributes*. In other words, the attributes that characterize the subgraphs of one class are different from the attributes around which the subgraphs from another class center upon. A stereotypical example to this insight is teenagers focusing on attributes such as 'selfies' and 'partying' whereas elderly being characterized by 'knitting' and 'tea partying'. Note that even though classes might share common attributes, we aim to identify those that are *exclusive* and not the overlapping ones; as those best help characterize the differences. Figure 1 presents a complete sketch of our problem.

A vast body of methods for community detection has been proposed for both simple [4, 8, 16, 26, 36] as well as attributed graphs

---

[*]Work performed while at Stony Brook University

---

[1]Here we focus on two classes for simplicity however our methods easily generalize to subgraphs from any number of classes.

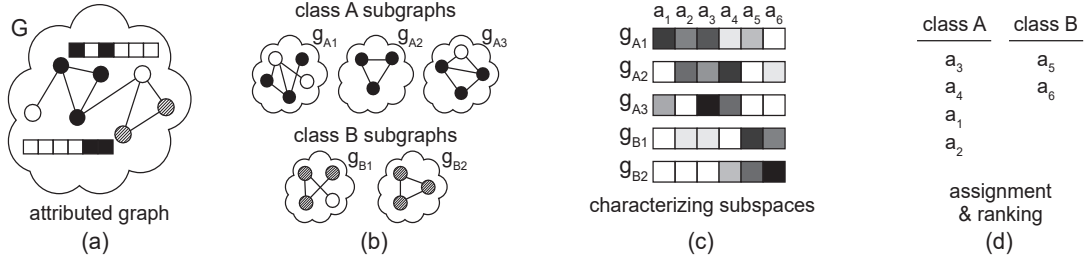

Figure 1: Problem sketch on toy data. Given (b) node-attributed subgraphs (or (a) nodes around which we extract subgraphs) from different classes (A and B), we find (c) the characterizing subspace (i.e., the focus attributes and respective weights) for each subgraph, and (d) split and rank the attributes for characterizing and comparing the classes.

[2, 12, 13, 15, 23, 25, 28, 33, 37]. Those are primarily concerned with extracting disjoint or overlapping groups of nodes, while optimizing some graph clustering objective. Our problem is considerably different. Unlike them, our goal is to understand the differences between distinct classes of subgraphs (or communities) through the attributes that characterize them, not to extract better ones.

Similar studies have been done in characterizing and comparing the social media use of different classes of subjects. For example, features from a user's social media interactions have been shown to predict demographic information such as gender [5], age [29, 31], occupation [30], location [9, 18], and income [10]. More nuanced traits have also been predicted about individuals, such as personality [32], or mental illness [6]. However, these studies tend to focus solely on text attributes and do not consider broader levels of social interactions in a network.

A recent work in the same lines with ours is by DellaPosta *et al.* [7], which studied *network effects* for explaining how political ideology becomes linked to lifestyles, such as "latte liberals" and "bird-hunting conservatives". Their simulated models reveal strong indications for influences operating *between* individuals in political "echo chambers" rather than within individuals, demonstrating evidence toward "one is the company they keep", i.e., that social ties matter.

In this work, we analyze the differences between individuals from different classes. Unlike previous work which has focused primarily on the individual's attributes (mostly text), we use local communities around individual nodes in addition to attributes to characterize them. Specifically, our contributions include the following:

- We introduce the general characterization problem for a given collection of attributed subgraphs from different classes—which leverages *both* the structure of social ties as well as the attributes. Our formulation entails partitioning the attributes into as many groups as the number of classes, while maximizing the total attributed quality score of the input subgraphs (§3).
- We show that our attribute-to-class assignment problem is NP-hard and an optimal $(1 - 1/e)$-approximation algorithm exists (§4.1). We also propose two different faster heuristics that are linear-time in the number of attributes and subgraphs (§4.2).
- Through extensive experiments, we compare the performance of the algorithms, present findings that agree with human intuition on real-world scenarios from 3 datasets, and demonstrate that our *characterization* approach is better suited to sense-making than *discriminative* classification approaches (§5).

## 2 PROBLEM DEFINITION

In this section, we introduce the notation used throughout text and present the formal problem definition. An attributed graph $G = (V, E, A)$ is a graph with node set $V$, undirected edges $E \subseteq V \times V$, and a set of attributes $A = \{a_1, \ldots a_d\}$ associated with every node, where $\mathbf{a_i} \in \mathfrak{R}^d$ denotes the $d$-dimensional attribute vector for node $i$. In this work we consider real-valued and binary attributes. A categorical attribute can be transformed to binary through one-hot encoding.

Given a collection of attributed subgraphs from $c$ classes, our aim is to *split* the attributes in $A$ into $c$ *disjoint* groups such that the total quality score $Q$ of all the subgraphs based on function $q(\cdot)$ and their assigned attributes is maximized. Here we use the normality measure [27] for $q(\cdot)$, which can be replaced with any other measure of interest that can utilize both graph structure and attributes in general.

Our problem statement is given for two classes as follows for simplicity, which can be generalized to multiple classes.

*Definition 2.1 (Characterization Problem).*
***Given***
- *$p$ attributed subgraphs $g_1^+, g_2^+, \ldots, g_p^+$ from class 1,*
- *$n$ attributed subgraphs $g_1^-, g_2^-, \ldots, g_n^-$ from class 2, from graph $G$, and attribute vector $\mathbf{a} \in \mathfrak{R}^d$ for each node;*

***Find***
- *a partitioning of attributes to classes as $A^+$ and $A^-$, where $A^+ \cup A^- = A$ and $A^+ \cap A^- = \emptyset$,*
- *focus attributes $A_i^+ \subseteq A^+$ (and respective weights $\mathbf{w}_i^+$) for each subgraph $g_i^+$, $\forall i$, and*
- *focus attributes $A_j^- \subseteq A^-$ (and respective weights $\mathbf{w}_j^-$) for each subgraph $g_j^-$, $\forall j$;*

***such that***
- *total quality $Q$ of all subgraphs is maximized, where $Q = \sum_{i=1}^{p} q(g_i^+|A^+) + \sum_{j=1}^{n} q(g_j^-|A^-)$;*

***Rank*** *attributes within $A^+$ and $A^-$.*

The above problem contains three subproblems, in particular, (P1) how to measure the quality of an attributed subgraph, (P2)

how to find the focus attributes (and their weights) of a given subgraph, and (P3) how to assign and rank the attributes for different classes so as to maximize total quality. In practice, classes focus on a small set of attributes. Further, our ranking of the attributes ensures those irrelevant to both classes and those common between them are ranked lower and only a few of the most differentiating attributes stand out. Figure 1 shows an example for our problem for 5 subgraphs from 2 classes, where 6 attributes are split into two and ranked for characterizing and comparing the classes.

In the next section, we address the subproblems in the given order above, in §3.1 through §3.3 respectively, to build up a solution for our main problem statement.

## 3 FORMULATION
### 3.1 Quantifying Quality

To infer the characterizing subspace for a given subgraph, we use a measure of subgraph quality. The idea is to find the attribute subspace and respective weights that maximize the quality of each subgraph. In this work, we use the normality measure [27], which not only utilizes both subgraph structure as well as attributes, but also quantifies both internal and external connectivity of the subgraph.

For a given subgraph $g$, its normality $N(g)$ is given as in Eq. (1), where $W$ is the adjacency matrix, $k_i$ is node $i$'s degree, $sim(\cdot)$ is the similarity function of attribute vectors weighted by $\mathbf{w_g}$, $e$ is the number of edges, and $B(g)$ denotes the nodes at the boundary of the subgraph (for isolated subgraphs, $B(g)$ is empty). The two terms in (1) respectively quantify $g$ internally and externally: many, surprising, and highly similar connections inside $g$ increase internal quality, whereas if such edges are at the boundary, they decrease external quality. For technical details of normality, see [27].

$$\begin{aligned}
N(g) = I + X &= \sum_{i \in g, j \in g} \left(W_{ij} - \frac{k_i k_j}{2e}\right) sim(\mathbf{a_i}, \mathbf{a_j} | \mathbf{w_g}) \\
&\quad - \sum_{\substack{i \in g, b \in B(g) \\ (i,b) \in E}} \left(1 - \min(1, \frac{k_i k_b}{2e})\right) sim(\mathbf{a_i}, \mathbf{a_b} | \mathbf{w_g}) \\
&= \mathbf{w_g}^T \cdot (\mathbf{a}_I + \mathbf{a}_X)
\end{aligned} \quad (1)$$

One can handle highly heterogeneous attributes simply by choosing the right $sim(\cdot)$ function. Also note that $\mathbf{a}_I$ and $\mathbf{a}_X$ are vectors that can be directly computed from data. Attributes with large non-zero weights in $\mathbf{w_g}$ are called the *focus attributes* of subgraph $g$.

### 3.2 Discovering Characterizing Subspaces

For a subgraph we can use Eq. (1) to compute its normality provided $\mathbf{w_g}$, the weights for the (focus) attributes. However the focus is often latent and hard to guess without prior knowledge, especially in high dimensions where nodes are associated with a long list of attributes. Even if the focus is known apriori, it is hard to manually assign weights.

Instead, we *infer* the attribute weight vector for a given subgraph, so as to maximize its normality score. In other words, we leverage normality as an objective function to infer the best $\mathbf{w_g}$ for a given $g$. This objective is written as

$$\begin{aligned}
&\max_{\mathbf{w_g}} \quad \mathbf{w_g}^T \cdot (\mathbf{a}_I + \mathbf{a}_X) \\
&\text{s.t.} \quad \|\mathbf{w_g}\|_p = 1, \ \mathbf{w_g}(a) \geq 0, \ \forall a = 1 \ldots d
\end{aligned} \quad (2)$$

Note that $\mathbf{w_g}$ is normalized to its $p$-norm to restrain the solution space. We also introduce non-negativity constraint on the weights to facilitate their interpretation. In the following we let $\hat{\mathbf{x}} = (\mathbf{a}_I + \mathbf{a}_X)$, where $\hat{\mathbf{x}}(a) \in [-1, 1]$.

If one uses $\|\mathbf{w_g}\|_{p=1}$, or the $L_1$ norm, the solution picks the *single* attribute with the largest $\hat{\mathbf{x}}$ entry as the focus. That is, $\mathbf{w_g}(a) = 1$ where $\max(\hat{\mathbf{x}}) = \hat{\mathbf{x}}(a)$ and 0 otherwise. This can be interpreted as the most important attribute that characterizes the subgraph. Note that $\hat{\mathbf{x}}$ may contain only negative entries, in which case the largest negative entry is selected, and the subgraph is deemed as low quality.

If there are *multiple* attributes that can increase normality, we can also select all the attributes with positive entries in $\hat{\mathbf{x}}$ as the subgraph focus. The weights of these attributes, however, should be proportional to the magnitude of their $\hat{\mathbf{x}}$ values. This is exactly what $\|\mathbf{w_g}\|_{p=2}$, or the $L_2$ norm yields. It is shown (see [27]) that under $p = 2$,

$$\mathbf{w_g}(a) = \frac{\hat{\mathbf{x}}(a)}{\sqrt{\sum_{\hat{\mathbf{x}}(i)>0} \hat{\mathbf{x}}(i)^2}} , \quad (3)$$

where $\hat{\mathbf{x}}(a) > 0$ and 0 otherwise, such that $\mathbf{w_g}$ is unit-normalized. The normality score of subgraph $g$ then becomes $N(g) = \mathbf{w_g}^T \cdot \hat{\mathbf{x}} = \sum_{\hat{\mathbf{x}}(a)>0} \frac{\hat{\mathbf{x}}(a)}{\sqrt{\sum_{\hat{\mathbf{x}}(i)>0} \hat{\mathbf{x}}(i)^2}} \hat{\mathbf{x}}(a) = \sqrt{\sum_{\hat{\mathbf{x}}(i)>0} \hat{\mathbf{x}}(i)^2} = \|\hat{\mathbf{x}}_+\|_2$, i.e., the 2-norm of $\hat{\mathbf{x}}$ induced on the attributes with positive $\hat{\mathbf{x}}$ entries.

### 3.3 Identifying Class Differences

*3.3.1 Splitting attributes between classes.* In this last part we return to our main problem statement, where we seek to *split* the attribute space between different classes so as to be able to identify their differences. We aim to obtain such an assignment of attributes with a goal to maximize the total quality (i.e., normality) of all the subgraphs from both classes. This ensures that the subgraphs are still characterized well, even under the constraint that the attributes are not shared across classes.

Let $S^+ = \{g_1^+, \ldots, g_p^+\}$ and $S^- = \{g_1^-, \ldots, g_n^-\}$ denote the sets of all subgraphs in class 1 and class 2, respectively, where each subgraph is now associated with a $d$-dimensional non-negative vector $\mathbf{x}$. This is the same as the $\hat{\mathbf{x}}$ vector introduced in §3.2, except that all the negative entries are set to zero. Recall that the entries of $\hat{\mathbf{x}}$ depict the contribution of each attribute to the quality of the subgraph. Therefore, we can drop the negative entries (recall that the optimization in (2) selects only the positive entries, if any).[2]

The goal is then to find two disjoint attribute groups $A^+$ and $A^-$, $A^+ \cup A^- = A$ and $A^+ \cap A^- = \emptyset$, such that the total quality of all subgraphs is maximized (see problem statement in §2). Given a set of selected attributes $S$, the quality of a subgraph $g$ can be written as

$$N(g|S) = \sqrt{\sum_{a \in S} \mathbf{x}(a)^2} = \|\mathbf{x}[S]\|_2 \quad (4)$$

i.e., the 2-norm of $\mathbf{x}$ induced on the attribute subspace. Therefore, the overall problem can be (re)formulated as

$$\max_{A^+ \subseteq A, A^- \subseteq A} \quad \frac{1}{p} \sum_{i \in \mathcal{S}^+} \|\mathbf{x}_i[A^+]\|_2 + \frac{1}{n} \sum_{j \in \mathcal{S}^-} \|\mathbf{x}_j[A^-]\|_2$$

$$\text{such that} \quad A^+ \cap A^- = \emptyset \quad (5)$$

---
[2]There may be subgraphs for which $\hat{\mathbf{x}}$ contains only negative entries. We exclude such subgraphs from the study of discovering class differences, as they are deemed low quality.

Note that we normalize the terms by the number of subgraphs in each class to handle class imbalance. We also emphasize that our objective in (5) is different from a classification problem in two key ways. First, we work with $\mathbf{x}$ vectors that embed information on subgraph connectivity as well as focus attributes rather than the original attribute vectors $\mathbf{a}$'s. Second, our objective embraces *characterization* and aims to find a partitioning of attributes that maximizes total quality, which is different from finding a decision boundary that minimizes classification loss as in *discriminative* approaches (See §5).

*3.3.2 Ranking attributes.* A solution to (5) (next section) provides a partitioning of the attributes into two groups. We can analyze the specific attributes assigned to classes to characterize their differences. Since this is an exploratory task, analyzing a large number of attributes would be infeasible. For easier interpretation, we need a ranking of the attributes.

One could think of using $\sum_{i \in \mathcal{S}^{(c)}} N(g_i | a \in A^{(c)})$ for scoring each attribute $a$. This however does not reflect the differentiating power but only the importance of $a$ for class $c$. We want both important and differentiating attributes to rank higher as they truly characterize the difference between subgraphs of the two classes. Specifically, some attributes may exhibit positive $\mathbf{x}$ entries for a particular class, however very small values, indicating only slight relevance. We may also have some attributes that exhibit large positive $\mathbf{x}$ entries, however for *both* classes. While relevant, such attributes are non-differentiating and would be uninformative for our task.

To get rid of only slightly relevant or non-differentiating attributes and obtain a sparse solution, we define a *relative contribution* score $rc(\cdot)$ for each attribute $a$ as

$$rc(a) = \frac{1}{p} \sum_{i \in \mathcal{S}^+} \mathbf{x}_i(a) - \frac{1}{n} \sum_{j \in \mathcal{S}^-} \mathbf{x}_j(a) \quad (6)$$

which is the difference between $a$'s contribution alone to the average quality of subgraphs in class 1 and class 2. We then rank the attributes within each class by their $rc$ values.

## 4 ALGORITHMS
### 4.1 Optimal Approximation

It is easy to show that our quality function $N(g|S) = \|\mathbf{x}[S]\|_2$ in Eq. (4) is a monotone submodular set function with respect to $S$ for non-negative $\mathbf{x}$. That is, the quality of a subgraph increases monotonically with increasing set size $S$. In addition, the increase follows the *diminishing returns* property known in economics, i.e., adding a new attribute $a$ to a set $S$ increases the function less than adding the same attribute to its smaller subset $S'$; $N(g|a \cup S) - N(g|S) \leq N(g|a \cup S') - N(g|S')$, $S' \subseteq S \subseteq A$.

Under this setting, we find that our problem in (5) can be stated as an instance of the Submodular Welfare Problem (SWP), which is defined as follows.

*Definition 4.1 (Submodular Welfare Problem).* Given $d$ items and $m$ players having monotone submodular utility functions $w_i : 2^{[d]} \to \mathfrak{R}_+$, find a partitioning of the $d$ items into $m$ disjoint sets $I_1, I_2, \ldots, I_m$ in order to maximize $\sum_{i=1}^{m} w_i(I_i)$.

In our formulation items map to the attributes for $d = |A|$, whereas players correspond to the classes, in the simplest case for $m = 2$. In addition, the utility function is written for each class $c \in \{+, -\}$ as

$$w_c(I_c) = N(\mathcal{S}^{(c)} | A^{(c)}) = \frac{1}{n^{(c)}} \sum_{k \in \mathcal{S}^{(c)}} \|\mathbf{x}_k[A^{(c)}]\|_2 \quad (7)$$

which is the average normality scores of subgraphs $\mathcal{S}$ belonging to class $c$. As $\| \cdot \|_2$ is a monotone and submodular function, so is $N(\mathcal{S}^{(c)})$ since the sum of submodular functions is also submodular [21]. Note that although we focus on two classes in this work, the SWP is defined more generally for $m$ players, i.e., classes. As such, it is easy to generalize our problem to more classes following the same solutions introduced for the SWP.

The SWP is first studied by Lehmann *et al.* [19], who proposed a simple on-line greedy algorithm that gives a 1/2-approximation for this problem. Later, Vondrák proposed an improved $(1 - 1/e)$-approximation solution [35]. Khot *et al.* showed that the SWP cannot be approximated to a factor better than $1 - 1/e$, unless P = NP [17]. Mirrokni *et al.* further proved that a better than $(1 - 1/e)$-approximation would require exponentially many value queries, regardless of P = NP [24]. As such, Vondrák's solution is the *optimal* approximation algorithm for the SWP, which we use to solve our problem in (5). The solution uses a multilinear extension to relax the subset optimization into a numerical optimization problem such that advanced optimization techniques, in particular a continuous greedy algorithm, can be applied. The continuous solution is then rounded to obtain a near-optimal set with the aforementioned guarantee [35].

### 4.2 Faster Heuristics

*4.2.1 Pre-normalized weights.* For the formulation shown in (5), we unit-normalize the attribute weights as in Eq. (3), only based on a selected subset $S$: $\mathbf{w_g}(a) = \frac{\mathbf{x}(a)}{\sqrt{\sum_{a \in S} \mathbf{x}(a)^2}}$. This normalization yields the quality function $N(g|S) = \sqrt{\sum_{a \in S} \mathbf{x}(a)^2}$, and requires that $S$ is given/known. A way to simplify this function is to *fix* the attribute weights at $\mathbf{w_g}(a) = \frac{\mathbf{x}(a)}{\sqrt{\sum_{a \in A} \mathbf{x}(a)^2}}$, i.e., to normalize them based on all the (known) positive attributes in $A$ rather than a (unknown) subset. This way the attribute weights can be *pre-computed* and do not depend on the to-be-selected attribute subsets. The simplified version of the maximization in (5) is then written as

$$\max_{A^+ \subseteq A, A^- \subseteq A} \quad \frac{1}{p} \sum_{i \in \mathcal{S}^+} \sum_{a \in A^+} \frac{\mathbf{x}_i(a)^2}{D_i} + \frac{1}{n} \sum_{j \in \mathcal{S}^-} \sum_{a \in A^-} \frac{\mathbf{x}_j(a)^2}{D_j}$$

$$\text{such that} \quad A^+ \cap A^- = \emptyset \quad (8)$$

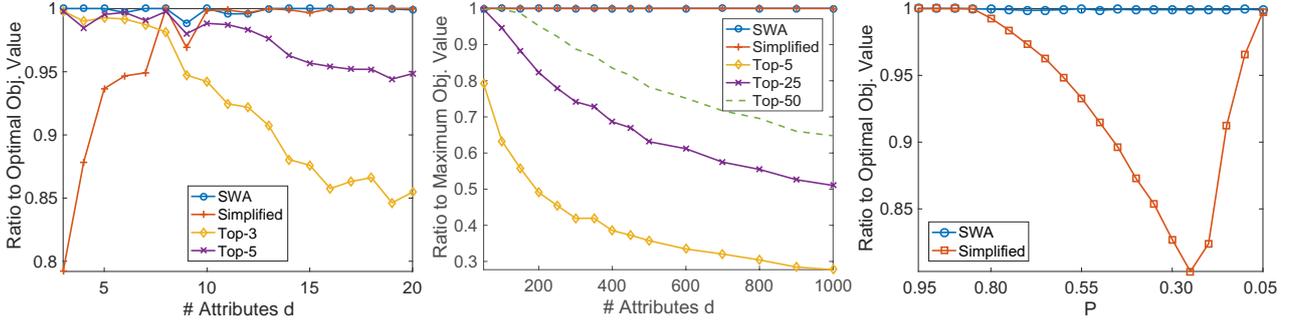

Figure 2: Ratio of objective value achieved by each test algorithm (left) to optimal value found by brute-force ($d = 3, 4, \ldots, 20$) and (center) to maximum achieved value ($d = 50, 100, \ldots, 1000$). (right) Performance of SIMPLIFIED degrades as it assigns nearly all attributes to the class with higher expected x values, ignoring the diminishing returns property of the objective function, whereas SWA remains near-optimal under all settings. All results are averaged over 10 random datasets.

where the denominator $D_i = \sqrt{\sum_{a \in A} \mathbf{x}_i(a)^2} = \|\mathbf{x}_i\|_2$, which can now be treated as constant as it does not depend on $A^+$ (same for $D_j$).

The simplified function $N(g|S) = \frac{\|\mathbf{x}[S]\|_2^2}{\|\mathbf{x}\|_2}$ is now a monotone *modular* function with respect to $S$. The contribution of a particular new attribute to the quality of a subgraph does not any more depend on the other attributes that are already in the selected set. That is, $N(g|a \cup S) - N(g|S) = N(g|a \cup S') - N(g|S') = \frac{\mathbf{x}(a)^2}{\|\mathbf{x}\|_2}, \forall S, S' \subseteq A$.

As a result, a simple linear-time algorithm can be employed to solve the objective in (8). The algorithm iterates over the attributes (order does not matter), and assigns each attribute $a$ to the class $c$ for which the average subgraph quality is improved more than others, that is, $\arg\max_c \frac{1}{n^{(c)}} \sum_{k \in S^{(c)}} \frac{\mathbf{x}_k(a)^2}{\|\mathbf{x}_k\|_2}$, breaking ties arbitrarily.

While the objective values of the solutions to (5) and (8) are likely to differ due to the difference in computing the attribute weights, the weight normalization does not change the order of the attributes by importance within a given set. Therefore, we conjecture that the two solutions will perform similarly, which we investigate through experiments in §5.

#### 4.2.2 Top-k attributes per class.
For exploratory tasks, such as understanding the class differences via characterizing attribute subspaces, it would be most interesting to look at the top $k$ most important attributes from each class. We also expect each subgraph to exhibit only a handful of focus attributes (experiments on real-world graphs confirm this intuition). Therefore, limiting the analysis to a top few attributes would be sufficient.

For a given (small) $k$, finding the top $k$ attributes $A_{*k}$ that maximize $N(g|A_{*k})$ for a *single* subgraph $g$ is easy—that would be the $k$ attributes with the largest values in $g$'s x (see Eq. (4)). However, we have a *multi-criterion* objective, with a goal to find the top $k$ attributes that maximize the total normality for all subgraphs from a class at once rather than a single one, that is $N(S^{(c)}|A_{*k}^{(c)})$ (see Eq. (7)).

The multi-criterion optimization problem is NP-hard [21]. On the other hand, we know that $\|\cdot\|_2$ is a monotone submodular function, and so is the class quality function $N(\cdot) = \sum \|\cdot\|_2$. As such, we find the top $k$ attributes for each class separately, using the lazy greedy hill-climbing algorithm introduced in [21]. Since these (separate) solutions may end up having common attributes, we *resolve* the solutions by assigning each common attribute only to the class for which its *average contribution* is higher (the individual terms in Eq. (6)). The search repeats until each class gets assigned $k$ unique attributes. Finally, $k$ is not a critical parameter to set, but rather can be chosen interactively.

## 5 EVALUATION

We evaluate our approach to the characterization problem and proposed algorithms on both synthetic and real-world datasets. Our goal is to answer the following questions:
- How do the proposed algorithms perform and compare to each other? What is their scalability and runtime?
- Are the findings on real-world data meaningful?
- How does characterization compare to classification?

### 5.1 Analysis on Synthetic Datasets

Through synthetic data experiments, our goal is to compare the algorithmic and computational performance of the proposed algorithms, respectively in terms of objective value achieved and running time. Specifically, we compare:
- SWA (Submodular Welfare Algorithm, §4.1)
- SIMPLIFIED (with pre-normalized weights, §4.2.1)
- TOP-$k$ (§4.2.2)

We generate the x vectors for $p = n = 100$ subgraphs each from $c = 2$ classes, while varying the number of attributes $d$. The $\mathbf{x}(a)$ values of each feature $a$ for subgraphs from class $c$ are drawn from a Normal distribution with a distinct $\mu_a^c$ and $\sigma_a^c$. The $\mu_a^c$'s themselves are drawn from a *zero-mean unit-variance* Normal (note that those attributes with negative mean tend to be less relevant for the class). The $\sigma_a^c$'s are randomly drawn from a $[0, 1]$ uniform distribution.

**Algorithmic performance.** In the first experiment, we test the *optimality* of the algorithms by comparing their achieved objective value to that of *brute-force* where we try all possible partitionings to identify the optimal solution. We experiment with $d = \{3, 4, \ldots, 20\}$ as brute-force is not computationally feasible for more than 20 attributes, and $k = \{3, 5\}$ for TOP-$k$.

In Figure 2 (left) we report the ratio between each test algorithm's objective value and the optimal as found by brute-force (ratio 1

means they are equal) with varying attribute size. The results are averaged over 10 random realizations of the synthetic datasets as described above. We notice that SWA achieves near-optimal performance throughout, which Simplified catches up with as the number of attributes $d$ increases. Top-$k$ loses optimality as $k$ becomes smaller compared to $d$, where the decline is faster for smaller $k$.

Figure 2 (center) is similar, where we compare performances under larger attribute sizes $d = \{50, 100, \ldots, 1000\}$. As $d$ is larger, we also use larger $k = \{5, 25, 50\}$. As brute-force cannot be computed in reasonable time, we report ratios w.r.t. the maximum objective value achieved among the tested algorithms. We find that Simplified achieves near-identical performance to SWA. Again, the ratios of Top-$k$ methods drop as the attribute space grows. Interestingly, Top-50 (out of 1000) attributes from each of both classes yield 64.78% of the maximum objective value.

While it may appear that Simplified performs as well as SWA, we can show that under certain conditions where the diminishing returns property of submodular problems plays a major role, it becomes inferior to SWA. To show such a setting, we design an experiment where the $\mathbf{x}(a)$ values of an attribute $a$ are drawn uniformly from $[P, 1]$ for class 1, and from $[0, 1 - P]$ for class 2 as we decrease $P$ from 0.95 to 0.05. Note that as the ranges (and hence the variance) of the values increase, the expected value of *every* attribute remains higher for class 1. The results are shown in Figure 2 (right). For large $P$, the values for class 1 are significantly larger and both algorithms assign all attributes to class 1. As the ranges start overlapping and the expected values get closer, Simplified continues to assign all attributes to class 1 (with higher expected value) even though the marginal increase to the objective value decreases significantly as we go on due to diminishing returns. As the variance gets even larger, Simplified again performs similar to SWA as it starts assigning some attributes to class 2 due to the random variation. Arguably, it is unlikely to encounter this setting in real-world datasets, where there exist many similarly-distributed attributes for sufficiently different classes of subgraphs.

**Computational performance.** Finally, we compare the proposed algorithms in terms of their running time and scalability, as the number of attributes grows. Figure 3 shows runtime in seconds for $d = \{50, 100, \ldots, 1000\}$. We note that all the algorithms scale near-linearly. SWA has the largest slope with increasing $d$, while finishing under 8 seconds for $d = 1000$ and $p = n = 100$ subgraphs from two classes. The scalability of Top-$k$ depends on $k$ which decreases with increasing $k$. Simplified heuristic lies in the bottom and is reliably one of our fastest methods.

Overall, SWA and Simplified work best on all datasets. Simplified can be parallelized easily, as each attribute is processed independently. For massive datasets, one can also fall back to Top-$k$, which is capable of identifying the few key attributes for characterization.

### 5.2 Analysis on Real-world Datasets

For real-world data analysis we consider attributed graphs where nodes are assigned class labels. We study the class differences of nodes by the "company that they keep". That is, we characterize each node with a local community surrounding them, using the local community extraction algorithm of Andersen *et al.* [3]. One can also use ego-networks, where a node is grouped with all its immediate neighbors.

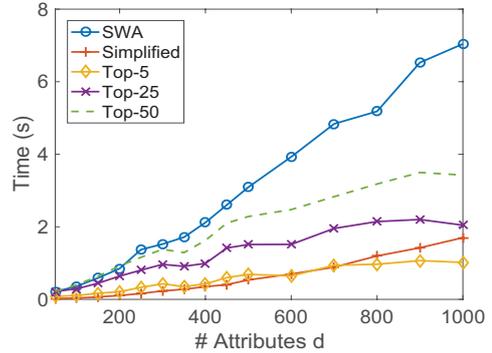

**Figure 3: Running time (seconds) with increasing number of attributes. All algorithms scale near-linearly with $d$.**

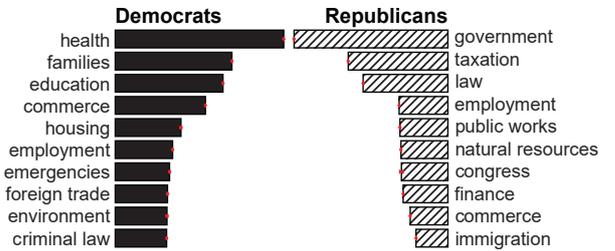

**Figure 4: Top 10 attributes in ranked order for Democrats and Republicans in `Congress`. Characterizing attributes reveal the contrast between liberal and conservative ideas.**

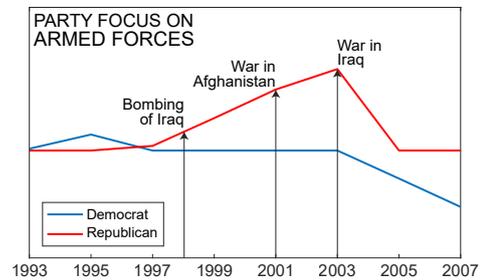

**Figure 5: Change of focus on attribute "National Security and Armed Forces" among Democrats and Republicans. We observe increased interest by Republicans in time of war.**

We report the top-10 attributes by *relative contribution* in (6) per class side by side for comparison. To be precise, we randomly sample 90% of our subgraphs 100 times and present the *average relative contribution* (bars) and standard deviation (error bars) so as to ensure that our results are not an artifact of the subgraphs at hand.

We experiment with 3 real-world attributed networks: (*i*) bill co-sponsorships of Congressmen [11], (*ii*) co-purchase network of

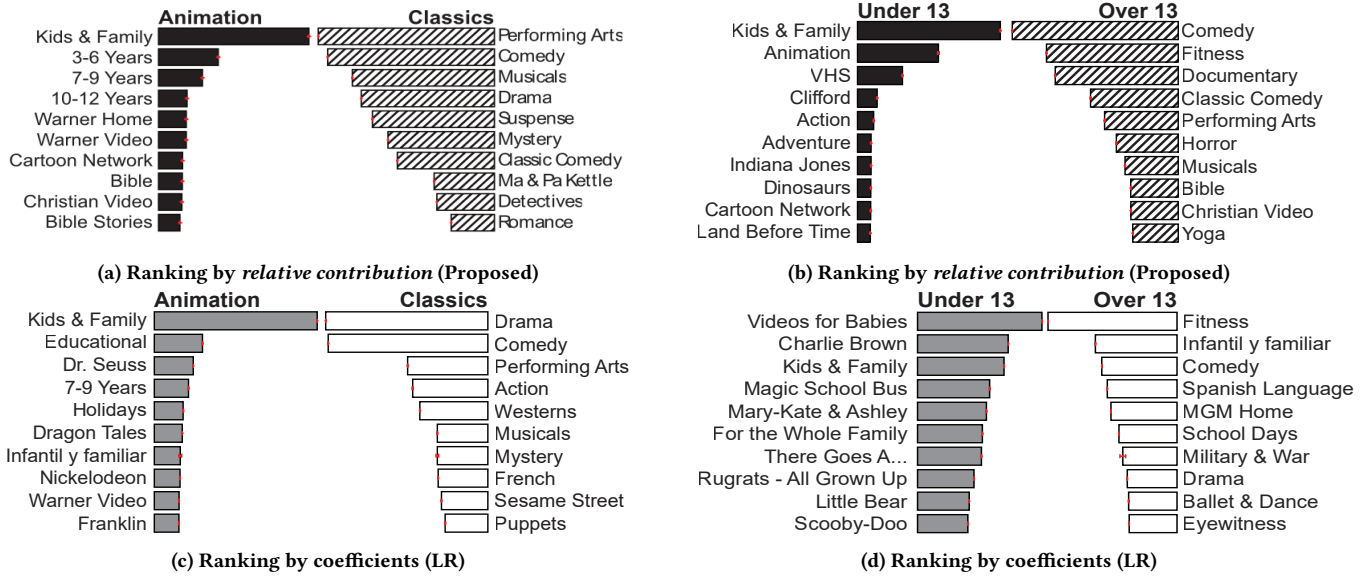

Figure 6: Characterization vs. Classification: Logistic Regression (LR) prefers infrequent attributes that discriminate well. Our proposed method discovers subspaces that characterize the data in a more natural way.

Amazon videos [20], and (*iii*) DBLP co-authorship network. We describe the individual datasets and present our findings next.

Congress. We consider 8 co-sponsorship networks from the 103rd Congress to 110th. The nodes are congressmen. An edge depicts co-sponsorship of a bill by two congressmen, and the edge weight is the number of times two nodes sponsored a bill together. Each bill is assigned a phrase that describes its subject, with a total of 32 such phrases. We mirror these bill subjects to their sponsors to create node attributes. The networks are highly dense, so we remove low-weighted edges such that the size of the giant connected component maintains more than 95% of its original size.

Figure 4 presents the top-ranking attributes among two classes, Democrats and Republicans (averaged over 8 congresses in the dataset). As expected, Democrats have a liberal agenda centered upon social and environmental programs, while Republicans mainly focus on regulating government, immigration and financial issues.

Since the Congress dataset is temporal, we can also explore how the focus of the two parties changes on a particular subject over time. A clear example of this is bills on "National Security and Armed Forces". Figure 5 shows the *average contribution* of this attribute for both parties (individual terms in Eq.(6)) over years. Starting with the US conflict with Iraq, this attribute seizes Republicans' attention, and it continues to grow after the 9/11 attacks and the beginning of the war in Afghanistan. It reaches its peak during the start of war in Iraq and then starts to lose attention towards the last years when the US troops are withdrawn from the middle-east. This abnormal change in interest in national security and armed forces is especially interesting since before and after the years of international crisis, this attribute has close to zero attention, even achieves negative values during the last years as a Democrat attribute, which indicates it is not characteristic of neither of the parties.

Amazon. This network contains 4011 nodes, 9487 edges, and 903 attributes. Nodes are videos from amazon.com, and edges depict co-purchase relations between two videos, indicating that they are frequently bought together. Attributes range from describing the video genre such as "Comedy" and "Drama", to the age-range of the audience intended for the videos such as "7-9 Years", popular franchises like "Sponge Bob Series", the form of the videos like "Animation", and the device it comes in such as "VHS" or "DVD".

We have experimented with two scenarios on Amazon to showcase the strength of our method in characterizing different classes. We set semantically different attributes as classes and use the rest for characterization. The specific queries are (1) *Animation* vs. *Classics* and (2) Videos for *Under 13* years old and Videos for *Over 13* years old. *Under 13* class consists of the videos exhibiting the attributes "Birth - 2 Years", "3 - 6 Years", "7 - 9 Years" and "10 - 12 Years". Rest of the videos belong to the *Over 13* class.

Figure 6a and 6b respectively show top-10 attributes per class as ranked by our method on these two scenarios. We find that "Kids & Family" and age groups "3-12 Years" are key characterizing attributes for *Animation*. "Warner Videos" and "Cartoon Network" are also among the top attributes. Perhaps surprisingly, "Christian" videos and "Bible" stories follow the above. On the other hand, we note genre-related attributes that truly define *Classics*, such as "Performing Arts", "Comedy", and "Musicals".

For the second scenario, we observe "Kids & Family" and "Animation" attributes to mostly characterize the *Under 13* videos. In contrast, the characterizing attributes for *Over 13* are those that cannot really be consumed by children, including "Comedy", "Fitness", and "Documentary" videos.

DBLP. This network contains 134K nodes, 1.478M edges and 2K attributes. Nodes are computer scientists and links are co-authorship relations. Attributes are computer science conferences and journals, where a node exhibits an attribute if s/he has at least one publication in the venue.

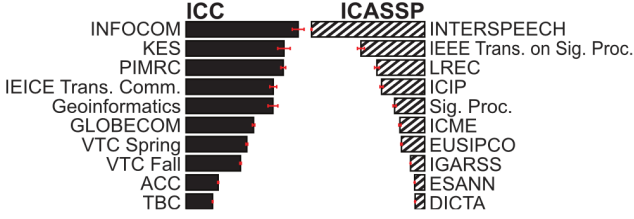

**Figure 7: Top 10 attributes in ranked order for *ICC* and *ICASSP* conferences in** DBLP.

The classes are *ICC*, a conference on communications, vs. *ICASSP*, a conference on speech and signal processing. We randomly sample 100 nodes from all nodes of each class and find subgraphs around them, to maintain a manageable set of subgraphs. Figure 7 shows top-10 attributes for the two classes. We find that attributes for *ICC* revolve around networking, communications and mobile technologies, including "INFOCOM", "GLOBECOM" and "PMIRC", while attributes for *ICASSP* are conferences on speech, video and image processing and linguistics, including "INTERSPEECH", "ICIP" and "EUSIPCO".

## 5.3 Characterization vs. Classification

Here we examine the differences between characterization and discriminative classification. Similar to our method, a sparse solution from a regularized classifier will contain a subspace of the attributes, which can be ranked using an attribute importance score from the model. Such regularized sparse methods are popular approaches for exploratory data analysis and form the foundations of many interpretable modeling methods.

For this comparison, we use Logistic Regression (LR) with LASSO regularization [34] to learn a sparse solution. We first train a LR model for binary classification between classes $c^+$ and $c^-$ using raw node attributes as input. After classification, we use the LR coefficients to partition the attributes between the classes, assigning those with positive coefficients to $c^+$ and negative ones to $c^-$, and rank by their magnitude. To overcome class imbalance in the dataset, we oversample from the small class to make the class sizes equal and then do the classification. We repeat this procedure 10 times to eliminate the effect of such sampling.

Figure 6 compares the top-ranking attributes obtained with our method to those found through LR. We see that LR prefers infrequent, but highly discriminating attributes. For example, consider the difference in attributes found between our method and LR for the *Animation* vs. *Classics* classification task (Fig. 6a and 6c). Here, LR completely fails to assign high weight to two attributes ("3-6 Years" and "10-12 Years") that are both very prevalent in the dataset. Instead its third strongest attribute is "Dr. Seuss", which is perfectly discriminative (all items with this attribute are *Animation*), but is present in only 4% of the nodes. This is a clear example of sacrificing characterization for discrimination.

We see this behavior again in Figure 6d, where LR ranks rare attributes highly (such as "Charlie Brown" and "Mary-Kate & Ashley") above more frequent attributes which are quite discriminative ("Kids & Family" and "Animation"). In contrast to LR, our proposed method ranks attributes by their contribution across subgraphs, finding a subspace of attributes which better characterizes the input subgraphs.

Intuitively, when an attribute is present in many of the nodes that belong to a class, it is considered to be a characterizing attribute of that class. On the other hand, when observing an attribute at a node indicates a high probability of the node belonging to a particular class, then the attribute is a discriminative one for that class. To quantify these, we use class support and confidence, metrics commonly used in association rule mining [1].

Let $\#(c, a)$ denote the number of nodes in class $c$ with attribute $a$, $\#(a)$ total number of nodes with attribute $a$, and $\#(c)$ total number of nodes from class $c$, then:

- **Confidence(C)**: probability of belonging to class $c$ when attribute $a$ is observed in some node: $Cfd(c, a) = Pr(c|a) = \frac{\#(c,a)}{\#(a)}$
- **Class Confidence(CC)**: probability of belonging *only* to class $c$, when attribute $a$ is observed in some node: $CC(c^+, a) = Pr(c^+|a) - Pr(c^-|a)$
- **Support(S)**: percentage of nodes in class $c$ that exhibit attribute $a$: $Sup(c, a) = \frac{\#(c,a)}{\#(c)}$
- **Class Support(CS)**: difference of support for $a$ between classes: $CS(c^+, a) = Sup(c^+, a) - Sup(c^-, a)$

As we seek distinct subspaces, we only use the relative metrics, i.e. *CC* and *CS*. Ideally having an attribute high on both metrics is best, however this case happens rarely. An attribute with high *class support* can be considered as a good representative of a class while an attribute with high *class confidence* can be used for classification purposes. To measure the average characterization of the attributes assigned to a given class, we have:

$$\overline{CS}(c, A^{(c)}) = \frac{\sum_{a \in A^{(c)}} w_a CS(c, a)}{\sum_{a \in A^{(c)}} w_a}, \quad (9)$$

where $\overline{CS}$ is the *weighted* average of CS over all attributes assigned to class $c$. Weight $w_a$ here is the metric that we use for ranking attributes in corresponding methods, which is the *relative contribution* for our proposed approach and the absolute coefficient values for LR. Likewise, to measure the total discrimination of a set of attributes, we have:

$$\overline{CC}(c, A^{(c)}) = \frac{\sum_{a \in A^{(c)}} w_a CC(c, a)}{\sum_{a \in A^{(c)}} w_a}, \quad (10)$$

where again, $\overline{CC}$ is the weighted average of CC for all attributes assigned to class $c$.

Figure 8 presents both measures for the two ranking methods for the real-world scenarios. Our method outperforms LR w.r.t. the characterization aspect ($\overline{CS}$). This is to be expected – as our method searches for attributes present in a focused subspace across many subgraphs of a class and ranks them accordingly. Surprisingly, in nearly all cases, the subspaces we find also have a comparable discriminative power ($\overline{CC}$) to LR. The *Under 13* case where we have low $\overline{CC}$ is when most discriminating attributes (as ranked high by LR) are infrequent and hence have low support. In fact, 8/10 of top attributes in Figure 6d have CS value < 0.1.

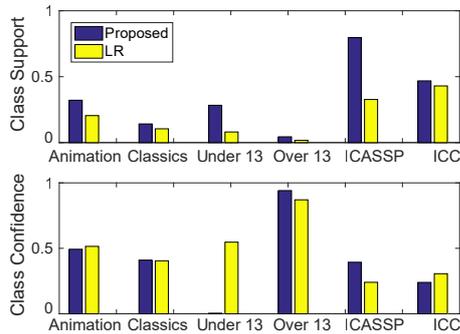

**Figure 8:** $\overline{CS}$ and $\overline{CC}$ for our method and LR. Our method always has superior *Average Class Support* when compared to LR, and also has competitive *Average Class Confidence*. (LR is explicitly optimizing for discrimination, so it is expected to do better in this regard)

## 6 CONCLUSION

Studies have shown evidence for characteristic differences between individuals of different genders, age groups, political orientations, personalities, etc. In this work, we generalized and mathematically formalized the characterization problem of attributed subgraphs from different classes. Our solution is through a lens into the node attributes as well as the social ties in their local networks. We showed that our problem, of partitioning attributes between classes so as to maximize the total quality of input subgraphs, is NP-hard, and that the proposed algorithms find near-optimal solutions and scale well with the number of attributes. Extensive experiments on synthetic and real-world datasets demonstrated the performance of the algorithms, the suitability of our approach for qualitative exploratory analysis, and its advantage over discriminating approaches.

## ACKNOWLEDGMENTS

This research is sponsored by NSF CAREER 1452425 and IIS 1408287, DARPA Transparent Computing Program under Contract No. FA8650-15-C-7561, ARO Young Investigator Program under Contract No. W911NF-14-1-0029, and a faculty gift from Facebook. Any conclusions expressed in this material are of the authors and do not necessarily reflect the views, expressed or implied, of the funding parties.